\documentclass[prc,aps,amsmath,showpacs,twocolumn,floatfix]{revtex4}
\usepackage{bm}
\usepackage{amsmath}
\usepackage{epsfig}
\newcommand{\Nh}{N^{-1}}
\newcommand{\nh}{n^{-1}}
\newcommand{\ph}{p^{-1}}

\newcommand{\bh}{b^{-1}}
\newcommand{\vk}{\bm{k}}
\newcommand{\vp}{\bm{p}}
\newcommand{\pfa}{p_F^{(a)}}
\newcommand{\pfb}{p_F^{(b)}}
\newcommand{\pfaq}{p_F^{(a)\,2}}
\newcommand{\pfbq}{p_F^{(b)\,2}}
\newcommand{\pfn}{p_F^{(n)}}
\newcommand{\pfp}{p_F^{(p)}}
\newcommand{\pfnq}{p_F^{(n)\,2}}
\newcommand{\pfpq}{p_F^{(p)\,2}}
\newcommand{\efn}{\epsilon_F^{(n)}}
\newcommand{\efp}{\epsilon_F^{(p)}}
\newcommand{\efe}{\epsilon_F^{(e)}}
\begin{document}
\title{Pionic susceptibility for charged pions in asymmetric nuclei}
\author{Michael Urban}
\affiliation{Institut de Physique Nucl{\'e}aire, F-91406 Orsay Cedex,
France}
\author{Jochen Wambach}
\affiliation{Institut f{\"u}r Kernphysik, Schlossgartenstr.\@ 9,
D-64289 Darmstadt, Germany}
\begin{abstract}
At low energies the particle-hole ($N\Nh$) part of the pionic susceptibility in
isospin-symmetric nuclear matter is known to behave very differently from the
susceptibility in finite nuclei due to the presence of an energy gap in the
$N\Nh$ excitation spectrum. In this note we show that for charged pions in
$N\neq Z$ nuclei the changes due to the gap are very similar to those in the
symmetric case, except at very low momenta, where a qualitatively different
behavior is found.
\end{abstract}
\maketitle1
In several applications \cite{Oset,Kohl} the polarization function of a
nuclear medium is needed for very low energy transfer $\omega \approx 0$. As
an example, we will consider here the pionic susceptibility $\chi(\omega,k)$
\cite{EricsonWeise}, but our discussion applies likewise to other isovector
channels. For finite nuclei, $\chi(\omega,k)$ is often calculated in a
local-density approximation using results from the homogenous medium.
In this case the particle-hole ($N\Nh$) contribution, which is dominant
at low energies, is described by the Lindhard function. However, when
applied to finite nuclei, this approximation is justified only if the energy
$\omega$ is much larger than the lowest-lying $N\Nh$ excitation energy. The
modifications due to this energy gap in the excitation spectrum were
discussed  in Ref.\@ \cite{Oset} for the isospin-symmetric case and are
summarized below. In this note we will focus on the isospin-asymmetric case.

For charged pions in isospin-asymmetric matter the original Lindhard function
\cite{Lindhard,FetterWalecka} cannot be used. Instead, it is convenient to
split the Lindhard function into direct and crossed contributions. We define
\begin{equation}
\phi_{a\bh}(\omega,k) = 2 \int\frac{d^3 p}{(2\pi)^3}\,
  \frac{\theta(\pfb-p)\theta(|\vp+\vk|-\pfa)}
    {\omega-\frac{k^2+2\vk\cdot\vp}{2m}+i\varepsilon}\,,
\label{defphi}
\end{equation}
where $a$ and $b$ indicate neutron ($n$) or proton ($p$), $\pfa$ and
$\pfb$ denote the corresponding Fermi momenta, and $m$ is the nucleon mass.
For completeness, the explicit expression for $\phi_{a\bh}$ is given in the
appendix. Using this definition, e.g., the non-interacting 
(Fermi-gas) susceptibility for a $\pi^+$ is
given by
\begin{equation}
\chi_{\pi^+}^{(0)}(\omega,k) = 2\frac{f^2}{m_\pi^2}
  \big[\phi_{p\nh}(\omega,k)+\phi_{n\ph}(-\omega,k)\big]\,,
\label{chiusual}
\end{equation}
where $f$ denotes the $\pi NN$ coupling constant. The susceptibility for a
$\pi^-$ can be obtained by replacing $\omega\leftrightarrow-\omega$. In
realistic cases one has to include corrections from the short-range $NN$
correlations (Ericson-Ericson Lorentz-Lorenz effect \cite{EricsonWeise})
which lead to an RPA resummation of $\chi_{\pi}^{(0)}$. 

In symmetric matter the indices $a$ and $b$ can be omitted. In this case it
is obvious from Eq.\@ (\ref{defphi}) that, for $\omega\neq 0$, the
function $\phi$ fulfills $\phi(\omega \neq 0, k\rightarrow 0) = 0$ as a
consequence of Pauli blocking, whereas for $\omega = 0$ one finds
$\phi(\omega = 0,k\rightarrow 0) = -m p_F/(2 \pi^2)$\,.
The reason for this behavior is that in the latter case both numerator
and denominator of Eq.\@ (\ref{defphi}) vanish. However, as pointed out in
Ref.\@ \cite{Oset}, this results in a bad approximation for the
susceptibility of a finite nucleus. Due to the finite excitation
energy of the lowest-lying $N\Nh$ excitation, the
denominator cannot vanish for $\omega = 0$. As a consequence
the $N\Nh$ part of the susceptibility of a finite nucleus must go to zero as
$k\rightarrow 0$. Following Ref.\@ \cite{Oset}, this effect is easily 
included by adding an energy gap $\Delta$ to the $N\Nh$ excitation energy
$(k^2+2\vk\cdot\vp)/(2m)$. Then the pionic susceptibility can be written as
\begin{equation}
\chi_\pi^{(0)}(\omega,k) = 2\frac{f^2}{m_\pi^2}
  \big[\phi(\omega-\Delta,k)+\phi(-\omega-\Delta,k)\big]\,.
\label{chiDelta}
\end{equation}
The drastic effect of this modification can be seen from the dashed lines
in Fig.\@ 1 which are in agreement with Fig.\@ 2 of Ref.\@ \cite{Oset}. The
short-dashed line corresponds to the usual Lindhard function, Eq.\@
(\ref{chiusual}), whereas the long-dashed line represents the Lindhard function
with a gap, Eq.\@ (\ref{chiDelta}).
\begin{figure}
{\centering\epsfig{file=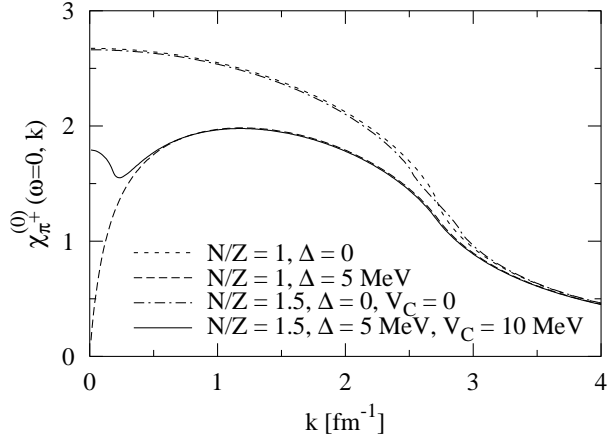,width=8cm}}
\caption{$N\Nh$ part of the pionic susceptibility $\chi_{\pi^+}^{(0)}$ at
nuclear matter density ($\varrho_n+\varrho_p = 0.17\,\mathrm{fm}^{-3}$) for
fixed $\omega = 0$ as function of $k$.
Dashed lines: symmetric matter, corresponding to
$\pfn = \pfp = 1.36\,\mathrm{fm}^{-1}$.
Short dashes: usual Lindhard function, Eq.\@ (\ref{chiusual}).
Long dashes: modified Lindhard function with gap in the $N\Nh$ excitation
spectrum, Eq.\@ (\ref{chiDelta}).
Dash-dotted and solid lines: asymmetric matter with
  $\varrho_n = 1.5 \varrho_p$,
corresponding to $\pfn = 1.45\,\mathrm{fm}^{-1}$ and
$\pfp = 1.26\,\mathrm{fm}^{-1}$.
Dash-dotted line: usual Lindhard function, generalized to the asymmetric case,
Eq.\@ (\ref{chiDelta}).
Solid line: with gap in the excitation spectrum and proton energies shifted
by $V_C$, adjusted such that $\efp = \efn$.}
\end{figure}

We now return to the asymmetric case. To be specific, we will
discuss the case of neutron-rich matter, i.e., $\pfn > \pfp$. In this case
$\phi_{p\nh}(\omega,k)$ is non-zero for all values of $\omega$ and $k$, since
Pauli blocking can never be complete. This is in contrast to
$\phi_{n\ph}(\omega,k)$, which is identically zero due to Pauli
blocking for all momenta $k\leq\pfn-\pfp$. It should also be noted that
$\phi_{p\nh}(\omega=0,k)$ and thus also $\chi^{(0)}_{\pi^+}(\omega=0,k)$
are complex for $k<2\pfn$.

The previous discussion is valid for infinite matter which is adequate
e.g., for the interior of a neutron star. In view of the strong modifications
in the symmetric case due to the energy gap in finite nuclei, it is
interesting to include similar corrections also in the asymmetric case. We
will consider two modifications:
\begin{itemize}
\item[1.]{A neutron star is stabilized against $\beta$ decay by the Fermi
energy of the electrons, $\efe = \efn-\efp$. The Coulomb potential is zero as
a consequence of charge neutrality. This situation is very different in finite
nuclei, where the whole Fermi sea of the protons is shifted upwards with
respect to the Fermi sea of the neutrons by the Coulomb potential $V_C$, such
that for $\beta$-stable nuclei the Fermi energies of protons and neutrons are
almost equal (see, e.g., Fig.\@ 1(a) of Ref.\@ \cite{Hamamoto}). Therefore we
shift all proton energies by $V_C\approx (\pfnq-\pfpq)/(2m)$\,.}
\item[2.]{As in the symmetric case, we include the energy gap $\Delta$ in
order to account for the discrete $N\Nh$ excitation spectrum at low energies.
It should be noted, however, that even if there is a gap in the $p\ph$ or
$n\nh$ excitation spectrum, there is no gap in the $p\nh$ spectrum of
$\beta$-unstable nuclei.}
\end{itemize}
Including both effects, the pionic susceptibility for a $\pi^+$ reads
\begin{multline}
\chi_{\pi^+}^{(0)}(\omega,k) = 2\frac{f^2}{m_\pi^2}
  \big[\phi_{p\nh}(\omega-\Delta-V_C,k)\\
  +\phi_{n\ph}(-\omega-\Delta+V_C,k)\big]\,.
\label{chiVCDelta}
\end{multline}
The result of these modifications concerns the imaginary part of
$\chi_{\pi^+}(\omega,k)$, which now vanishes for $\omega = 0$. The real part
changes in a way very similar to the symmetric case, as can be seen from
Fig.\@ 1, where the dashed-dotted line corresponds to the susceptibility of
infinite asymmetric matter, Eq.\@ (\ref{chiusual}), while the solid line
contains the energy gap $\Delta$ and the Coulomb potential $V_C$, Eq.\@
(\ref{chiVCDelta}). In fact, over a wide range of momenta it seems to be a
very accurate approximation to neglect the asymmetry completely. However, at
momenta below $\approx \pfn-\pfp$ the susceptibility stays almost constant
in the asymmetric case (solid line), while it goes to
zero for $k\rightarrow 0$ in the symmetric case (long-dashed line). As
mentioned above, this qualitative difference stems from the fact that in the
asymmetric case Pauli blocking can never be complete.
\begin{acknowledgments}
We thank A.\@ Richter for the suggestion to write this note. One of us
(M.U.) acknowledges support from the Alexander von Humboldt foundation.
\end{acknowledgments}
\appendix*
\section{Explicit expressions}
In the explicit expression for the function $\phi_{a\bh}(\omega,k)$ defined
in Eq.\@ (\ref{defphi}) four different cases must be distinguished:
\begin{itemize}
\item[(a)]{$k\geq \pfa+\pfb$ (no Pauli blocking): 
\begin{equation}
\phi_{a\bh}(\omega,k) = f^{(u)}_{a\bh}(\omega,k,-\pfb,\pfb)\,,
\label{phiunblocked}
\end{equation}}
\item[(b)]{$|\pfa-\pfb|\leq k<\pfa+\pfb$:
\begin{align}
\phi_{a\bh}(\omega,k) =
  &f^{(r)}_{a\bh}(\omega,k,\mbox{$\frac{\pfaq-\pfbq-k^2}{2k}$},\pfa-k)
    \nonumber\\
  &+f^{(u)}_{a\bh}(\omega,k,\pfa-k,\pfb)\,,
\end{align}}
\item[(c)]{$k<|\pfa-\pfb|$ and $\pfa\leq\pfb$:
\begin{align}
\phi_{a\bh}(\omega,k) =
  &f^{(u)}_{a\bh}(\omega,k,-\pfb,-\pfa-k)\nonumber\\
  &+f^{(r)}_{a\bh}(\omega,k,-\pfa-k,\pfa-k)\nonumber\\
  &+f^{(u)}_{a\bh}(\omega,k,\pfa-k,\pfb)\,,
\end{align}}
\item[(d)]{$k<|\pfa-\pfb|$ and $\pfa>\pfb$ (complete Pauli blocking):
\begin{equation}
\phi_{a\bh}(\omega,k) = 0\,.
\end{equation}}
\end{itemize}
In these equations $f^{(u)}_{a\bh}$ and $f^{(r)}_{a\bh}$ denote the
integrals over $p_\parallel$ and $p_\perp$ in the regions of $p_\parallel$
where the integration over $p_\perp$ is unrestricted or restricted by Pauli
blocking, respectively:
\begin{multline}
f^{(u)}_{a\bh}(\omega,k,p_1,p_2) = \frac{m}{16 \pi^2 k^3}\\
  \times\big[2 k(p_1-p_2)(k^2-k(p_1+p_2)-2m\omega)\\
  +\big(4 k^2\pfbq-(k^2-2m\omega^2)^2\big)\ln x\big]\,,
\end{multline}
\begin{multline}
f^{(r)}_{a\bh}(\omega,k,p_1,p_2) = \frac{m}{4\pi^2 k}
  \big[2 k(p_1-p_2)\\
  +(\pfbq-\pfaq+2m\omega)\ln x\big]\,,
\end{multline}
with
\begin{equation}
x = \frac{k^2+2 k p_1-2 m \omega -i\varepsilon}
  {k^2+2 k p_2-2m\omega-i\varepsilon}\,.
\end{equation}

If one is only interested in the real part of Eq.\@ (\ref{chiusual}), the
Pauli-blocking effects in the two terms cancel, and it is possible
to use the simple formula (\ref{phiunblocked}) for all cases, see Eqs.\@ (10.6)
to (10.8) of Ref.\@ \cite{Migdal}. However, for Eqs.\@ (\ref{chiDelta}) and
(\ref{chiVCDelta}), the distinction of the four cases (a) to (d) is necessary.

\end{document}